\documentclass[aps,floatfix,nofootinbib,showpacs,twocolumn,superscriptaddress]{revtex4}
\usepackage{graphicx}
\usepackage{dcolumn}
\usepackage{amsmath}
\usepackage{longtable}

\newcommand{\sfrac}[2]{\mbox{\footnotesize $\displaystyle \frac{#1}{#2}$}}
\newcommand{\bea}{\begin{eqnarray}}
\newcommand{\eea}{\end{eqnarray}}

\usepackage{color}
\definecolor{purple}{rgb}{0.5,0,0.5}
\definecolor{blue}{rgb}{0.0,0,0.9}
\usepackage[colorlinks=true, pdfstartview=FitV, linkcolor=purple, citecolor= purple, urlcolor=blue]{hyperref}

\begin{document}
\title{
Abelian anomaly and neutral pion production
}

\author{H.\,L.\,L.~Roberts}
\affiliation{Physics Division, Argonne National Laboratory, Argonne,
Illinois 60439, USA}
\affiliation{Institut f\"ur Kernphysik, Forschungszentrum J\"ulich, D-52425 J\"ulich, Germany}

\author{C.\,D.~Roberts}
\affiliation{Physics Division, Argonne National Laboratory, Argonne, Illinois 60439, USA}
\affiliation{Institut f\"ur Kernphysik, Forschungszentrum J\"ulich, D-52425 J\"ulich, Germany}
\affiliation{Department of Physics, Peking University, Beijing 100871, China}

\author{A.~Bashir}
\affiliation{Instituto de F\'{\i}sica y Matem\'aticas,
Universidad Michoacana de San Nicol\'as de Hidalgo, Apartado Postal
2-82, Morelia, Michoac\'an 58040, Mexico}

\author{L.\,X.~Guti\'errez-Guerrero}
\affiliation{Instituto de F\'{\i}sica y Matem\'aticas,
Universidad Michoacana de San Nicol\'as de Hidalgo, Apartado Postal
2-82, Morelia, Michoac\'an 58040, Mexico}

\author{P.\,C.~Tandy} \affiliation{Center for Nuclear Research, Department of
Physics, Kent State University, Kent OH 44242, USA}

\begin{abstract}
We show that in fully-self-consistent treatments of the pion; namely, its static properties and elastic and transition form factors, the asymptotic limit of the product $Q^2 G_{\gamma^\ast\gamma \pi^0}(Q^2)$, determined \emph{a priori} by the interaction employed, is not exceeded at any finite value of spacelike momentum transfer.
Furthermore, in such a treatment of a vector-vector contact-interaction one obtains a $\gamma^\ast\gamma \to \pi^0$ transition form factor that disagrees markedly with all available data.
We explain that the contact interaction produces a pion distribution amplitude which is flat and nonvanishing at the endpoints.  This amplitude characterises a pointlike pion bound-state.  Such a state has the hardest possible form factors; i.e., form factors which become constant at large momentum transfers and hence are in striking disagreement with completed experiments.
On the other hand, interactions with QCD-like behaviour produce soft pions, a valence-quark distribution amplitude that vanishes as $\sim (1-x)^2$ for $x\sim 1$, and results that agree with the bulk of existing data.
Our analysis supports a view that the large-$Q^2$ data obtained by the BaBar Collaboration is not an accurate measure of the $\gamma^\ast\gamma \to \pi^0$ form factor.
\end{abstract}
\maketitle

\section{Introduction}
The process $\gamma^\ast \gamma \to \pi^0$ is fascinating because in order to explain the associated transition form factor within the Standard Model on the full domain of momentum transfer, one must combine, using a single internally-consistent framework, an explanation of the essentially nonperturbative Abelian anomaly with the features of perturbative QCD.  The case for attempting this has received a significant boost with the publication of data from the BaBar Collaboration \cite{Aubert:2009mc} because, while they agree with earlier experiments on their common domain of squared-momentum-transfer \cite{Behrend:1990sr,Gronberg:1997fj}, the BaBar data are unexpectedly far \emph{above} the prediction of perturbative QCD at larger values of $Q^2$.

Herein we contribute toward understanding the discrepancy by analysing this process using the Dyson-Schwinger equations (DSEs) \cite{Roberts:1994dr,Roberts:2000aa,Maris:2003vk,%
Fischer:2006ub,RodriguezQuintero:2010wy,Roberts:2007jh}, which are known to have the capacity to connect nonperturbative and perturbative phenomena in QCD.  In particular, the connection between dynamical chiral symmetry breaking (DCSB) and
the Abelian \cite{Roberts:1987xc,Bando:1993qy,Roberts:1994hh,Alkofer:1995jx,Maris:1998hc,%
Bistrovic:1999dy,Holl:2005vu}
and non-Abelian \cite{Bhagwat:2007ha} anomalies is understood, as is the manner through which the perturbative QCD results for the large-$Q^2$ behaviour of transition form factor can be obtained \cite{Kekez:1998rw,Roberts:1998gs}.

As part of this analysis, we will elucidate the sensitivity of the $\gamma^\ast \gamma \to \pi^0$ transition form factor, $G_{\gamma^\ast \gamma \pi}(Q^2)$, to the pointwise behaviour of the interaction between quarks.  We will use existing DSE calculations \cite{Maris:2002mz} of this and the kindred $\gamma^\ast \gamma^\ast \to \pi^0$ form factor to characterise the $Q^2$-dependence of $G_{\gamma^\ast \gamma \pi}(Q^2)$ which is produced by a quark-quark interaction that is mediated by massless vector-bosons.  For comparison, we will compute the behaviour obtained if quarks interact instead through a contact interaction.  
Such comparisons are important to achieving a goal of charting the long-range behaviour of the strong-interaction in the Standard Model \cite{Aznauryan:2009da}.

In Sec.\,\ref{sec:model} we describe a symmetry-preserving regularisation and DSE-formulation of the contact interaction, and explain how a dressed-quark comes simultaneously to have a nonzero charge radius and a hard form factor.  In Sec.\,\ref{sec:TFF} we discuss the $\gamma^\ast \gamma \to \pi^0$ transition form factor in detail, describing: its connection with the Abelian anomaly; its asymptotic behaviour in QCD cf.\ that produced by a contact interaction; and how the nature of the interaction determines the pion's distribution amplitude.  Section~\ref{sec:reflect} places our results in context with extant data; and Sec.\,\ref{sec:end} expresses our conclusions.

\section{Bound state pion}
\label{sec:model}
\subsection{Bethe-Salpeter and gap equations}
Poincar\'e covariance entails that the Bethe-Salpeter amplitude for an isovector pseudoscalar bound-state of a dressed-quark and -antiquark takes the form
\begin{eqnarray}
\nonumber
\lefteqn{\Gamma_{\pi}^j(k;P) = \tau^{j}\gamma_5 \left[ i E_\pi(k;P) + \gamma\cdot P F_\pi(k;P) \right.}\\
&& \left. + \,  \gamma\cdot k \, G_\pi(k;P) + \sigma_{\mu\nu} k_\mu P_\nu H_\pi(k;P) \right],
\label{genpibsa}
\end{eqnarray}
where $k$ is the relative and $P$ the total momentum of the constituents, and $\{\tau^j,j=1,2,3\}$ are the Pauli matrices.\footnote{We employ a Euclidean metric with:  $\{\gamma_\mu,\gamma_\nu\} = 2\delta_{\mu\nu}$; $\gamma_\mu^\dagger = \gamma_\mu$; $\gamma_5= \gamma_4\gamma_1\gamma_2\gamma_3$; and $a \cdot b = \sum_{i=1}^4 a_i b_i$.  A timelike four-vector, $Q$, has $Q^2<0$.  Furthemore, we consider the isospin-symmetric limit.} This amplitude is determined from the homogeneous Bethe-Salpeter equation (BSE):
\begin{equation}
[\Gamma_{\pi}^j(k;P)]_{tu} = \int \! \frac{d^4q}{(2\pi)^4} [\chi_{\pi}^j(q;P)]_{sr} K_{tu}^{rs}(q,k;P)\,,
\label{genbse}
\end{equation}
where
%
%
$\chi_\pi^j(q;P) = S(q+P)\Gamma_{\pi}^j(q;P)S(q)$, $r,s,t,u$ represent colour, flavour and spinor indices, and $K$ is the quark-antiquark scattering kernel.  In Eq.\,(\ref{genbse}), $S$ is the dressed-quark propagator; viz., the solution of the gap equation:
\begin{eqnarray}
\nonumber \lefteqn{S(p)^{-1}= i\gamma\cdot p + m}\\
&&+ \int \! \frac{d^4q}{(2\pi)^4} g^2 D_{\mu\nu}(p-q) \frac{\lambda^a}{2}\gamma_\mu S(q) \frac{\lambda^a}{2}\Gamma_\nu(q,p) ,
\label{gendse}
\end{eqnarray}
wherein $m$ is the Lagrangian current-quark mass, $D_{\mu\nu}$ is the gluon propagator and $\Gamma_\nu$ is the quark-gluon vertex.

\subsection{Momentum-independent vector-boson exchange}
Following Ref.\,\cite{GutierrezGuerrero:2010md} we define
\begin{equation}
\label{njlgluon}
g^2 D_{\mu \nu}(p-q) = \delta_{\mu \nu} \frac{1}{m_G^2}\,,
\end{equation}
where $m_G$ is a gluon mass-scale (such a scale is generated dynamically in QCD, with a value $\sim 0.5\,$GeV \cite{Bowman:2004jm}) and proceed by embedding this interaction in a rainbow-ladder truncation of the DSEs.  This means $\Gamma_{\nu}(p,q) =\gamma_{\nu}$ in both Eq.\,(\ref{gendse}) and the construction of $K$ in Eq. (\ref{genbse}).  Rainbow-ladder is the leading-order in a nonperturbative, symmetry-preserving truncation \cite{Munczek:1994zz,Bender:1996bb}.  It is known and understood to be an accurate truncation for pseudoscalar mesons \cite{Bhagwat:2004hn,Chang:2009zb}.

With this interaction the gap equation becomes
\begin{equation}
 S^{-1}(p) =  i \gamma \cdot p + m +  \frac{4}{3}\frac{1}{m_G^2} \int\!\frac{d^4 q}{(2\pi)^4} \,
\gamma_{\mu} \, S(q) \, \gamma_{\mu}\,.   \label{gap-1}
\end{equation}
The integral possesses a quadratic divergence, even in the chiral limit.  If the divergence is regularised in a Poincar\'e covariant manner, then the solution is
\begin{equation}
\label{genS}
S(p)^{-1} = i \gamma\cdot p + M\,,
\end{equation}
where $M$ is momentum-independent and determined by
\begin{equation}
M = m + \frac{M}{3\pi^2 m_G^2} \int_0^\infty \!ds \, s\, \frac{1}{s+M^2}\,.
\end{equation}

To proceed we must specify a regularisation procedure.  We write \cite{Ebert:1996vx}
\begin{eqnarray}
\nonumber
\frac{1}{s+M^2} & = & \int_0^\infty d\tau\,{\rm e}^{-\tau (s+M^2)} \\
& \rightarrow & \int_{\tau_{\rm uv}^2}^{\tau_{\rm ir}^2} d\tau\,{\rm e}^{-\tau (s+M^2)}
\label{RegC}\\
& & =
\frac{{\rm e}^{- (s+M^2)\tau_{\rm uv}^2}-e^{-(s+M^2) \tau_{\rm ir}^2}}{s+M^2} \,, \label{ExplicitRS}
\end{eqnarray}
where $\tau_{\rm ir,uv}$ are, respectively, infrared and ultraviolet regulators.  It is apparent from Eq.\,(\ref{ExplicitRS}) that a nonzero value of $\tau_{\rm ir}=:1/\Lambda_{\rm ir}$ implements confinement by ensuring the absence of quark production thresholds \cite{Krein:1990sf,Roberts:2007ji}.  Furthermore, since Eq.\,(\ref{njlgluon}) does not define a renormalisable theory,  $\Lambda_{\rm uv}:=1/\tau_{\rm uv}$ cannot be removed but instead plays a dynamical role and sets the scale of all dimensioned quantities.

The gap equation can now be written
\begin{equation}
M = m +  \frac{M}{3\pi^2 m_G^2} \,{\cal C}^{\rm iu}(M^2)\,,
\end{equation}
where ${\cal C}^{\rm iu}(M^2)/M^2 = \Gamma(-1,M^2 \tau_{\rm uv}^2) - \Gamma(-1,M^2 \tau_{\rm ir}^2)$, with $\Gamma(\alpha,y)$ being the incomplete gamma-function.

Using the interaction we've specified, the homogeneous BSE for the pseudoscalar meson is ($q_+=q+P)$
\begin{equation}
\Gamma_{\pi}(P) = - \frac{4}{3} \frac{1}{m_G^2} \,
\int\frac{d^4 q}{(2\pi)^4} \, \gamma_{\mu} \chi_\pi(q_+,q) \gamma_{\mu}  \,. \label{Gamma-eq}
\end{equation}
With a symmetry-preserving regularisation of the interaction in Eq.\,(\ref{njlgluon}), the Bethe-Salpeter amplitude cannot depend on relative momentum.  Hence Eq.\,(\ref{genpibsa}) reduces to
\begin{equation}
\label{genpibsacontact}
\Gamma_\pi(P) = \gamma_5 \left[ i E_\pi(P) + \frac{1}{M} \gamma\cdot P F_\pi(P) \right].
\end{equation}
Crucially, $F_\pi(P)$, a component of pseudovector origin, remains.  It is an essential component of the pion, which has very significant measurable consequences and thus cannot be neglected.

\subsection{Ward-Takahashi identity}
Preserving the vector and axial-vector Ward-Takahashi identities is essential when computing properties of the pion.  The $m=0$ axial-vector identity states 
\begin{equation}
\label{avwti}
P_\mu \Gamma_{5\mu}(k_+,k) = S^{-1}(k_+) i \gamma_5 + i \gamma_5 S^{-1}(k)\,,
\end{equation}
where $\Gamma_{5\mu}(k_+,k)$ is the axial-vector vertex, which is determined by
\begin{equation}
\Gamma_{5\mu}(k_+,k) =\gamma_5 \gamma_\mu
- \frac{4}{3}\frac{1}{m_G^2} \int\frac{d^4q}{(2\pi)^4} \, \gamma_\alpha \chi_{5\mu}(q_+,q) \gamma_\alpha\,. \label{aveqn}
\end{equation}
To achieve this, one must implement a regularisation that maintains Eq.\,(\ref{avwti}).  To see what this entails, contract Eq.\,(\ref{aveqn}) with $P_\mu$ and use Eq.\,(\ref{avwti}) within the integrand.  This yields the following two chiral limit identities:
\begin{eqnarray}
\label{Mavwti}
M & = & \frac{8}{3}\frac{M}{m_g^2} \int\! \frac{d^4q}{(2\pi)^4} \left[ \frac{1}{q^2+M^2} +  \frac{1}{q_+^2+M^2}\right],\\
\label{0avwti}
0 & = & \int\! \frac{d^4q}{(2\pi)^4} \left[ \frac{P\cdot q_+}{q_+^2+M^2} -  \frac{P\cdot q}{q^2+M^2}\right]\,,
\end{eqnarray}
which must be satisfied after regularisation.  Analysing the integrands using a Feynman parametrisation, one arrives at the follow identities for $P^2=0=m$:
\begin{eqnarray}
 M & = &  \frac{16}{3}\frac{M}{m_G^2} \int\! \frac{d^4q}{(2\pi)^4} \frac{1}{[q^2+M^2]}, \label{avwtiMc}\\
0 & = & \int\! \frac{d^4q}{(2\pi)^4} \frac{\frac{1}{2} q^2 + M^2 }{[q^2+M^2]^2} \label{avwtiAc}.
\end{eqnarray}

Equation\,(\ref{avwtiMc}) is just the chiral-limit gap equation.  Hence it requires nothing new of the regularisation scheme.   On the other hand, Eq.\,(\ref{avwtiAc}) states that the axial-vector Ward-Takahashi identity is satisfied if, and only if, the model is regularised so as to ensure there are no quadratic or logarithmic divergences.  Unsurprisingly, these are the just the circumstances under which a shift in integration variables is permitted, an operation required in order to prove Eq.\,(\ref{avwti}).

We observe in addition that Eq.\,(\ref{avwti}) is valid for arbitrary $P$.  In fact its corollary, Eq.\,(\ref{Mavwti}), can be used to demonstrate that in the chiral limit the two-flavour scalar-meson rainbow-ladder truncation of the contact-interaction DSEs produces a bound-state with mass $m_\sigma = 2 \,M$ \cite{Roberts:2010gh}.  The second corollary, Eq.\,(\ref{0avwti}) entails
\begin{equation}
0 = \int_0^1d\alpha \,
\left[ {\cal C}^{\rm iu}(\omega(M^2,\alpha,P^2))  + \, {\cal C}^{\rm iu}_1(\omega(M^2,\alpha,P^2)\right], \label{avwtiP}
\end{equation}
with $\omega(M^2,\alpha,P^2) = M^2 + \alpha(1-\alpha) P^2$ and ${\cal C}^{\rm iu}_1(z) = - z (d/dz){\cal C}^{iu}(z)$.

\subsection{Pion's Bethe-Salpeter kernel}
\label{sec:pionkernel}
We are now in a position to write the explicit form of Eq.\,(\ref{Gamma-eq}):
\begin{equation}
\label{bsefinal0}
\left[
\begin{array}{c}
E_\pi(P)\\
F_\pi(P)
\end{array}
\right]
= \frac{1}{3\pi^2 m_G^2}
\left[
\begin{array}{cc}
{\cal K}_{EE} & {\cal K}_{EF} \\
{\cal K}_{FE} & {\cal K}_{FF}
\end{array}\right]
\left[\begin{array}{c}
E_\pi(P)\\
F_\pi(P)
\end{array}
\right],
\end{equation}
where
\begin{eqnarray}
\nonumber
{\cal K}_{EE} &= &\int_0^1d\alpha \left[ {\cal C}^{\rm iu}(\omega(M^2,\alpha,-m_\pi^2))\right.\\
&& \left.  + 2 \alpha(1-\alpha) \, m_\pi^2 \, \overline{\cal C}^{\rm iu}_1(\omega(M^2,\alpha,-m_\pi^2))\right],\\
{\cal K}_{EF} &=& -m_\pi^2 \int_0^1d\alpha\, \overline{\cal C}^{\rm iu}_1(\omega(M^2,\alpha,-m_\pi^2)), \\
{\cal K}_{FE} &=& \frac{1}{2} M^2 \int_0^1d\alpha\, \overline{\cal C}^{\rm iu}_1(\omega(M^2,\alpha,-m_\pi^2)),\\
{\cal K}_{FF} &=& - 2 {\cal K}_{FE}\,,
\end{eqnarray}
with $\overline{\cal C}_1(z) = {\cal C}_1(z)/z$.  We used Eq.\,(\ref{avwtiP}) to arrive at this form of ${\cal K}_{FF}$.

In the computation of observables, one must use the canonically-normalised Bethe-Salpeter amplitude; i.e., $\Gamma_\pi$ is rescaled so that
\begin{equation}
P_\mu = N_c\, {\rm tr} \int\! \frac{d^4q}{(2\pi)^4}\Gamma_\pi(-P)
 \frac{\partial}{\partial P_\mu} S(q+P) \, \Gamma_\pi(P)\, S(q)\,. \label{Ndef}
\end{equation}
In the chiral limit, this means
\begin{equation}
1 = \frac{N_c}{4\pi^2} \frac{1}{M^2} \, {\cal C}_1(M^2;\tau_{\rm ir}^2,\tau_{\rm uv}^2)
E_\pi [ E_\pi - 2 F_\pi].
\label{Norm0}
\end{equation}

With the parameter values (in GeV) \cite{GutierrezGuerrero:2010md}
\begin{equation}
\label{params}
m_G = 0.11 \,,\;
\Lambda_{\rm ir} = 0.24\,,\;
\Lambda_{\rm uv} = 0.823\,,
\end{equation}
one obtains the results presented in Table~\ref{resultsIR}.  We note that the leptonic decay constant and in-pion condensate are given by the following expressions \cite{Maris:1997hd,Maris:1997tm}:
\begin{eqnarray}
\label{fpim}
f_\pi & = & \frac{1}{M}\frac{3}{2\pi^2} \,[ E_\pi - 2 F_\pi] \,{\cal K}_{FE}^{P^2=-m_\pi^2} \,,\\
\label{kpim}
\kappa_\pi & = &  f_\pi \frac{3}{4\pi^2} [ E_\pi {\cal K}_{EE}^{P^2=-m_\pi^2} + F_\pi \,{\cal K}_{EF}^{P^2=-m_\pi^2} ]\,.
\end{eqnarray}
In the chiral limit $\kappa_\pi \to \kappa_\pi^0 = -\langle \bar q q \rangle$; i.e., the so-called vacuum quark condensate \cite{Brodsky:2010xf}.  Moreover, also in this limit, one may readily verify that \cite{GutierrezGuerrero:2010md}
\begin{equation}
\label{GTE}
E_\pi \stackrel{m=0}{=} \frac{M}{f_\pi}\,,
\end{equation}
which is a particular case of one of the Goldberger-Treiman relations proved in Ref.\,\cite{Maris:1997hd}.

\begin{table}[t]
\caption{Results calculated with the parameter values in Eq.\,(\ref{params}): $m_\pi$ is obtained from Eq.\,(\protect\ref{bsefinal0}); $\kappa_\pi$, $f_\pi$ are defined in Eqs.\,(\protect\ref{kpim}), (\protect\ref{fpim}); $m_\rho$ is determined by solving Eq.\,(\protect\ref{rhobse}); and the charge radii are discussed in connection with Eq.\,(\protect\ref{quarkradius}).
The static properties are commensurate with results from QCD-based DSE studies \protect\cite{Maris:1997tm}.
(Dimensioned quantities are listed in GeV or fm, as appropriate.)
\label{resultsIR}
}
\begin{center}
\begin{tabular*}
{\hsize}
{
l@{\extracolsep{0ptplus1fil}}
|c@{\extracolsep{0ptplus1fil}}
c@{\extracolsep{0ptplus1fil}}
|c@{\extracolsep{0ptplus1fil}}
l@{\extracolsep{0ptplus1fil}}
c@{\extracolsep{0ptplus1fil}}
c@{\extracolsep{0ptplus1fil}}
c@{\extracolsep{0ptplus1fil}}
|c@{\extracolsep{0ptplus1fil}}
c@{\extracolsep{0ptplus1fil}}
c@{\extracolsep{0ptplus1fil}}
}
 & $E_\pi$ & $F_\pi$ & M & $m_\pi$ & $^3\!\!\!\!\sqrt{\kappa_\pi}$ &$f_\pi$ &  $m_\rho$ & $r_q$ & $r_\pi$ & $r_\pi^\rho$
 \\\hline
m=0 & 4.28 & 0.69 & 0.40 & 0 & 0.22 & 0.094 & 0.90 & 0.34 & 0.30 & 0.45\\
m=0.008 & 4.36 & 0.72 & 0.41 & 0.14 & 0.22 & 0.094 & 0.91 & 0.33 & 0.30 & 0.44 \\
\end{tabular*} 
\end{center}
\end{table}

\subsection{Dressed-photon-quark vertex}
In coupling photons to a bound-state constituted from dressed-quarks, it is important that the quark-photon vertex be dressed so that it satisfy the vector Ward-Takahashi identity \cite{Roberts:1994hh}.  Indeed, where possible it should be dressed at a level consistent with the truncation used to compute the bound-state's Bethe-Salpeter amplitude \cite{Maris:1999bh}.  With our treatment of the interaction described in connection with Eq.\,(\ref{njlgluon}), the bare vertex $\gamma_\mu$ is sufficient to satisfy the Ward-Takahashi identity and ensure, e.g., a unit value for the charged pion's electromagnetic form factor \cite{GutierrezGuerrero:2010md}.  However, given the simplicity of the DSE kernels, one can readily do better.

A vertex dressed consistently with our rainbow-ladder pion is determined by the following inhomogeneous Bethe-Salpeter equation:
\begin{equation}
\label{GammaQeq}
\Gamma_\mu(Q) = \gamma_\mu - \frac{4}{3} \frac{1}{m_G^2} \int \frac{d^4 q}{(2\pi)^4} \, \gamma_\alpha \chi_\mu(q_+,q) \gamma_\alpha\,,
\end{equation}
where $\chi_\mu(q_+,q) = S(q+P) \Gamma_\mu (Q) S(q)$.  Owing to the momentum-independent nature of the interaction kernel, the general form of the solution is
\begin{equation}
\label{GammaQ}
\Gamma_\mu(Q) = \gamma^T_\mu P_T(Q^2) + \gamma_\mu^L P_L(Q^2)\,,
\end{equation}
where $Q_\mu \gamma^T_\mu = 0$ and $\gamma^T_\mu+\gamma^L_\mu=\gamma_\mu$.  This simplicity doesn't survive with a more sophisticated interaction.

Upon insertion of Eq.\,(\ref{GammaQ}) into Eq.\,(\ref{GammaQeq}), one can readily obtain
\begin{equation}
\label{PL0}
P_L(Q^2)= 1\,,
\end{equation}
owing to Eq.\,(\ref{0avwti}).  Using this same identity, one finds
\begin{equation}
\label{PTQ2}
P_T(Q^2)= \frac{1}{1+K_\gamma(Q^2)}
\end{equation}
with
\begin{eqnarray}
\nonumber
\lefteqn{K_\gamma(Q^2) = \frac{1}{3\pi^2m_G^2}}\\
& & \times \int_0^1d\alpha\, \alpha(1-\alpha) Q^2\,  \overline{\cal C}^{iu}_1(\omega(M^2,\alpha,Q^2))\,.
\end{eqnarray}
It is plain that
\begin{equation}
\label{PT0}
P_T(Q^2=0)=1\,,
\end{equation}
so that at $Q^2=0$ in the rainbow-ladder treatment of the interaction in Eq.\,(\ref{njlgluon}) the dressed-quark-photon vertex is equal to the bare vertex.\footnote{Equations~(\ref{PL0}), (\ref{PT0}) guarantee a massless photon and demonstrate that our regularisation also ensures preservation of the Ward-Takahashi identity for the photon vacuum polarisation \protect\cite{Burden:1991uh}.}
However, this is not true for $Q^2\neq 0$.  Indeed, the transverse part of the dressed-quark-photon vertex will exhibit a pole at that $Q^2<0$ for which
\begin{equation}
\label{rhobse}
1+K_\gamma(Q^2)=0\,.
\end{equation}
This is just the model's Bethe-Salpeter equation for the ground-state vector meson.  The mass obtained therefrom is listed in Table~\ref{resultsIR}.

\begin{figure}[t] 
\centerline{\includegraphics[clip,width=0.45\textwidth]{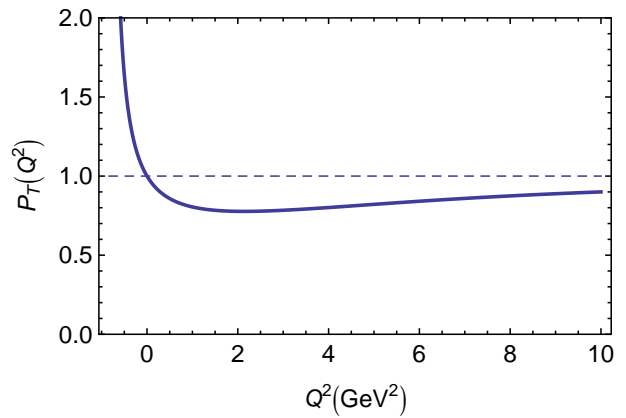}}
\caption{\label{qgammavertex} Dressing function for the transverse piece of the quark-photon vertex; viz., $P_T(Q^2)$ in Eq.\,(\protect\ref{PTQ2}).
}
\end{figure}

In Fig.\,\ref{qgammavertex} we depict the function that dresses the transverse part of the quark-photon vertex.  The pole associated with the ground-state vector meson is clear.  Another important feature is the behaviour at large spacelike-$Q^2$; namely,  $P_T(Q^2) \to 1^-$ as $Q^2\to \infty$.  This is the statement that a dressed-quark is pointlike to a large-$Q^2$ probe.  The same is true in QCD, up to the logarithmic corrections which are characteristic of an asymptotically free theory \cite{Maris:1999bh}.

One can define an electromagnetic radius for a dressed-quark; viz.,
\begin{equation}
\label{quarkradius}
r_q^2 = - 6 \left. \frac{d}{dQ^2} P_T(Q^2) \right|_{Q^2=0}.
\end{equation}
Our computed value is reported in Table~\ref{resultsIR}.  It is noteworthy that $r_q m_\rho/\sqrt{6}=0.63$ so that, although the ground-state vector-meson pole is a dominant feature of $P_T(Q^2)$ in the vicinity of $Q^2=0$, it does not completely determine the electromagnetic radius of the dressed-quark.

Nor, in fact, of anything else, as one can infer from the computed values of the pion charge radius reported in Table~\ref{resultsIR}:\footnote{In computing $r_\pi$, we follow Ref.\,\protect\cite{GutierrezGuerrero:2010md}.  Owing to the vector Ward-Takahashi identity, the longitudinal part of the vertex does not contribute to the pion's elastic form factor.} $r_\pi$ is obtained with $\Gamma_\mu(Q) = \gamma_\mu$; $r_\pi^\rho$ is computed with $\Gamma_\mu(Q) = \gamma_\mu P_T(Q^2)$; and $(r_\pi^\rho)^2 = r_\pi^2+r_q^2$, which is just a consequence of the product rule.  This emphasises again that single-pole vector-meson-dominance is a helpful phenomenology but not a hard truth \cite{Roberts:2000aa,Maris:2003vk,Maris:1999bh}.
%

\begin{figure}[t] 
\centerline{\includegraphics[clip,width=0.45\textwidth]{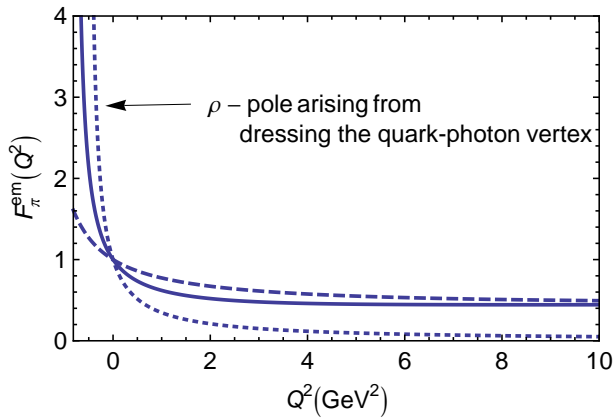}}
\caption{\label{figrhoFpi} $F^{\rm em}_{\pi}(Q^2)$ computed in rainbow-ladder truncation from the interaction in Eq.\,(\protect\ref{njlgluon}): \emph{solid curve} -- fully consistent, i.e., with a dressed-quark-photon vertex so that the $\rho$-pole appears; and \emph{dashed curve} -- computed using a bare quark-photon vertex.  \emph{Dotted curve} -- fit to the result in Ref.\,\protect\cite{Maris:2000sk}, which also included a consistently-dressed quark-photon vertex and serves to illustrate the trend of contemporary data.
}
\end{figure}

We show in Fig.\,\ref{figrhoFpi} that dressing the quark-photon vertex does not qualitatively alter the behaviour of $F^{\rm em}_\pi(Q^2)$ at spacelike momenta.  In particular, it does not change the fact that a momentum-independent interaction, Eq.\,(\ref{njlgluon}), regularised in a symmetry-preserving manner, produces\footnote{The rainbow-ladder truncation omits a so-called meson-cloud component of $F^{\rm em}_\pi(Q^2)$ but this, too, affects only the behaviour in a measurable neighbourhood of $Q^2=0$ \protect\cite{Alkofer:1993gu}.}
\begin{equation}
\label{Fpiconstant}
F_\pi^{\rm em}(Q^2 \to\infty) =\,{\rm constant.}
\end{equation}

\section{Transition form factor: \mbox{\boldmath $\gamma^\ast \pi^0 \gamma$}}
\label{sec:TFF}
In the rainbow-ladder truncation this process is computed from \cite{Holl:2005vu}
\begin{equation}
{\cal T}_{\mu\nu}(k_1,k_2) = T_{\mu\nu}(k_1,k_2)+ T_{\nu\mu}(k_2,k_1)\,,
\end{equation}
where the pion's momentum $P=k_1+k_2$, $k_1$ and $k_2$ are the photon momenta and
\begin{eqnarray}
\lefteqn{
T_{\mu\nu}(k_1,k_2) = \frac{\alpha_{\rm em}}{\pi f_\pi}\, \epsilon_{\mu\nu\alpha\beta}k_{1\alpha} k_{2\beta}\, G(k_1^2,k_1\cdot k_2,k_2^2)}\\
\nonumber
& = & \! {\rm tr}\!\!\! \int\!\!\!\frac{d^4 \ell}{(2\pi)^4} \, \chi_\pi(\ell_1,\ell_2) \,i {\cal Q}\Gamma_\mu(\ell_2,\ell_{12}) \, S(\ell_{12}) \, i {\cal Q} \Gamma_\nu(\ell_{12},\ell_1),\\
&& \label{anomalytriangle}
\end{eqnarray}
with $\ell_{1}=\ell-k_1$, $\ell_{2} = \ell + k_2$, $\ell_{12}=\ell - k_1 + k_2$, and ${\cal Q} = {\rm diag}[e_u,e_d] = e\, {\rm diag}[2/3,-1/3]$, $\alpha_{\rm em}=e^2/(4\pi)$.  The kinematic constraints are:
\begin{equation}
k_1^2=Q^2\,,\; k_2^2=0\,, \; 2\, k_1\cdot k_2=- (m_\pi^2+Q^2)\,.
\end{equation}

\subsection{Anomaly}
We first consider the chiral limit and $Q^2=0$, in which case  Eq.\,(\ref{anomalytriangle}) describes the ``triangle diagram'' that produces the Abelian anomaly and one must compute $G(0,0,0)$.  We have explained above that our regularisation of the interaction in Eq.\,(\ref{njlgluon}) ensures that the non-anomalous vector and axial-vector Ward-Takahashi identities are satisfied.  The outcome for the anomalous case is therefore very interesting.

Two contributions are obtained upon inserting Eq.\,(\ref{genpibsacontact}) into Eq.(\ref{anomalytriangle}); viz., one associated with $E_\pi(P)$, which we'll denominate $G_{E}$, and the other with $F_\pi(P)$, to be called $G_{F}$.  We first examine the latter.  To obtain $G_F(0,0,0)$ one need only expand the integrand in Eq.\,(\ref{anomalytriangle}) around $k_1=0=k_2$ and keep the term linear in $F_\pi(P) k_{1\alpha} k_{2\beta}$, a process which yields
\begin{eqnarray}
\nonumber\lefteqn{
G_F(0,0,0) = - \frac{f_\pi}{M}\int_0^\infty \!\! ds\,s^2\,F_\pi(P) \, \sigma_V(s) }\\
&& \rule{-1em}{0ex} \times \left[\sigma_V(s)^2 + s \sigma_V(s) \sigma_V^\prime(s) + \sigma_S(s)\sigma_S^\prime(s)\right], \label{GF000}
\end{eqnarray}
where $\sigma^\prime(s) = \frac{d}{ds}\sigma(s)$ and we have written
\begin{equation}
S(\ell) = -i \gamma\cdot \ell \, \sigma_V(\ell^2) + \sigma_S(\ell^2)\,.
\end{equation}
Using Eq.\,(\ref{genS}), one readily finds that $\sigma_V^\prime = -\sigma_V^2$, $\sigma_S^\prime = -M\sigma_V^2$.  These identities, when inserted into Eq.\,(\ref{GF000}), reveal that the integrand is identically zero, so that
\begin{equation}
G_F(0,0,0) =0\,.
\end{equation}
This is a particular case of the general result proved in Ref.\,\cite{Maris:1998hc}.  As explained therein, since the integral in Eq.\,(\ref{GF000}) is logarithmically divergent, the result is only transparent with the choice of momentum-partitioning that we have employed.

The remaining contribution is $G_E$, which, following the methods of Sec.\,\ref{sec:model}, can be written
\begin{equation}
G_E(0,0,0) = \frac{Mf_\pi}{\pi^2}\int d^4 \ell\,E_\pi(P) \sigma_V(\ell_{12}^2)\sigma_V(\ell_1^2)\sigma_V(\ell_2^2)\,.
\end{equation}
The integral is convergent and therefore a shift in integration variables cannot affect the result.  It follows that
\begin{equation}
\label{GE000open}
G_E(0,0,0) = E_\pi(P) \frac{f_\pi}{M} \int_0^\infty\!\! ds\, s  \frac{M^2}{(s+M^2)^3}\,.
\end{equation}

If we employ Eq.\,(\ref{RegC}), as with all other computations hitherto, this becomes
\begin{equation}
\label{GEreg}
G_E(0,0,0) =  \frac{1}{M^2} \, {\cal C}_2^{\rm iu}(M^2)\,,
\end{equation}
where ${\cal C}_2^{\rm iu}(z) = (z^2/2) (d^2/dz^2) {\cal C}^{\rm iu}(z)$ and we have used the Goldberger-Treiman relation in Eq.\,(\ref{GTE}).

In what has long been a textbook result, the anomalous Ward-Takahashi identity states that $G_E(0,0,0) = \frac{1}{2}$: truly, just this simple fraction.  Equation~(\ref{GEreg}) is plainly inconsistent with this because it produces a number that depends on the values of the parameters $\Lambda_{\rm ir}$, $\Lambda_{\rm uv}$.  Indeed, with the values in Eq.\,(\ref{params}), our regularisation of Eq.\,(\ref{njlgluon}) gives $f_\pi G_E(0,0,0) = 0.36$.  What has gone wrong?

The answer lies in the observation that
\begin{equation}
{\cal C}_2^{\infty\,0}(M^2) =
{\cal C}_2(M^2,\tau_{\rm ir}^2\to \infty ,\tau_{\rm uv}^2\to 0) = \frac{1}{2} \,M^2\,.
\end{equation}
One could have judged at the outset that no regularisation scheme which bounds the loop-integral can supply the correct result for the anomalous Ward-Takahashi identity because it blocks the crucial connection between the anomaly, topology and DCSB \cite{Witten:1983tw}.

To elucidate, return to Eq.\,(\ref{GE000open}).  The integral is convergent and dimensionless.  Hence, it cannot depend on $M$.  In a particular application of the procedure elucidated in Refs.\,\cite{Roberts:1987xc,Roberts:1994hh,Alkofer:1995jx,Maris:1998hc}, the change of variables $C(s) = M/s$ yields
\begin{eqnarray}
G_E(0,0,0) & = & \int_0^\infty\!\! dC\,\frac{1}{(1+C)^3} \\
& = & -\frac{1}{2} \int_0^\infty dC \frac{d}{dC}\frac{1}{(1+C)^2} =  \frac{1}{2}\,.
\label{halfanomaly}
\end{eqnarray}
The last line emphasises the connection between the simple rational-number result and the spacetime boundary: the anomaly is determined by the integral of a total derivative.  The result in Eq.\,(\ref{halfanomaly}) is obtained if, and only if, chiral symmetry is dynamically broken, since in this instance alone can Eq.\,(\ref{GTE}) be used to completely eliminate the pion structure-factor: $E_\pi(P)$, from the expression.

\subsection{Asymptotic behaviour}
\subsubsection{Massless vector-boson exchange}
\label{UVlimits}
In Ref.\,\cite{Lepage:1980fj}, using the methods of light-front quantum field theory, it was shown that
\begin{equation}
\label{BLuv}
\lim_{Q^2\to\infty} Q^2 G(Q^2,-\sfrac{1}{2}Q^2,0) = 4\pi^2 f_\pi^2.
\end{equation}
It is notable that this is a factor of $\pi/\alpha_s(Q^2)$ bigger than the kindred limit of the elastic pion form factor \cite{Lepage:1980fj,Farrar:1979aw,Efremov:1979qk};
i.e., at $Q^2=4\,$GeV$^2$, more than an order-of-magnitude larger.

Our analysis of Eq.\,(\ref{anomalytriangle}) is performed within a Poincar\'e covariant formulation.  In this case, as elucidated in Ref.\,\cite{Maris:2002mz}, the asymptotic limit of the doubly-off-shell process ($\gamma^\ast \gamma^\ast \to \pi^0$) is reliably computable in the rainbow-ladder truncation because both quark-legs in the dressed-quark-photon are sampled at the large momentum-scale $Q^2$, with the result \cite{Kekez:1998rw,Roberts:1998gs}
\begin{equation}
\label{gstgstuv}
\lim_{Q^2\to \infty} Q^2 G(Q^2,-Q^2-m_\pi^2/2,Q^2)
= \frac{2}{3} \, 4\pi^2 f_\pi^2,
\end{equation}
if the propagator of the exchanged vector-boson behaves as $1/k^2$ for large-$k^2$.  In order to obtain this result it is crucial that the pion's Bethe-Salpeter amplitude depends on the magnitude of the relative momentum and behaves as $1/k^2$ for large-$k^2$, as it does in QCD. (See also Sec.\,\ref{sec:pionDA}.)

Equations (\ref{BLuv}) and (\ref{gstgstuv}) correspond to the asymptotic limits of different but related processes.  Part of the mismatch owes to the fact that in the process $\gamma^\ast \gamma \pi^0$ not all quark-legs attached to vertices carry the large momentum-scale $Q^2$; namely, $\ell_2^2$ in Eq.\,(\ref{anomalytriangle}), and hence some amount of vertex dressing contributes, even at large $Q^2$.  This is consistent with a more general observation; namely, that in a covariant calculation any number of loops contribute to $\gamma^\ast \gamma \pi^0$ at leading order \cite{Lepage:1980fj}, and these provide a series of logarithmic corrections which should be summed.  The same is true of the pion's elastic form factor \cite{Maris:1998hc}.  Nevertheless, the correct power-law behaviour is necessarily produced.

\subsubsection{Contact interaction}
With the QCD-based expectation made clear, we now turn to the outcome produced by the contact interaction, Eq.\,(\ref{njlgluon}).  In this instance the arguments used to obtain Eq.\,(\ref{gstgstuv}) fail conspicuously because the pion's Bethe-Salpeter amplitude is completely independent of the relative momentum: all values of the relative momenta are equally likely.  This is why the interaction yields Eq.\,(\ref{Fpiconstant}), a result in striking disagreement with experiment.  A similar result is obtained in the present context.  However, a decision must be made before that can be exhibited.

Recall Eqs.\,(\ref{GF000}) and (\ref{GE000open}): the first is logarithmically divergent while the second is convergent even if the regularisation parameters are removed.  Indeed, one needs to remove the regularisation scales if the anomaly value is to be recovered.  However, the form factor is then ill-defined because the $F_\pi(P)$-term contributes the logarithmic divergence just noted.  We proceed by removing the regularisation in computing $G_E$ but retaining it in calculating $G_F$.  Notably, as we will see, with Eq.\,(\ref{njlgluon}) no internally consistent scheme can provide QCD-like ultraviolet behaviour but this prescription serves to preserve the infrared behaviour.

There is one more step in implementing this scheme.  In arriving at expressions such as those defining the pion's Bethe-Salpeter kernel (see Sec.\,\ref{sec:pionkernel}), we re-express a product of propagator-denominators via a Feynman parametrisation, then perform a change-of-variables, and thereafter rewrite the result using Eq.\,(\ref{RegC}).  This does not introduce any difficulties when the boundary at spacetime-infinity has no physical impact.
However, as we have seen, that is not the case with the anomaly.  The integral which defines $G_F$ is logarithmically divergent.  A shift of integration variables changes its value, and in doing that one runs afoul of the fact that it is impossible to simultaneously preserve the vector and axial vector Ward-Takahashi identities for triangle diagrams in field theories with axial currents that are bilinear in fermion fields.  Any shift in variables from that used in Eq.\,(\ref{anomalytriangle}) changes the value of $G_F(0,0,0)$.  We compensate by an additional regularising subtraction; i.e., by redefining
\begin{eqnarray}
\nonumber && G_F(Q^2,-(m_\pi^2+Q^2)/2,0) \\
& \rightarrow & G_F(Q^2,-(m_\pi^2+Q^2)/2,0) - G_F(0,0,0)\,.
\end{eqnarray}
In doing so we implement an anomaly-free electromagnetic current \cite{Jackiw}.

\begin{figure}[t] 
\centerline{\includegraphics[clip,width=0.45\textwidth]{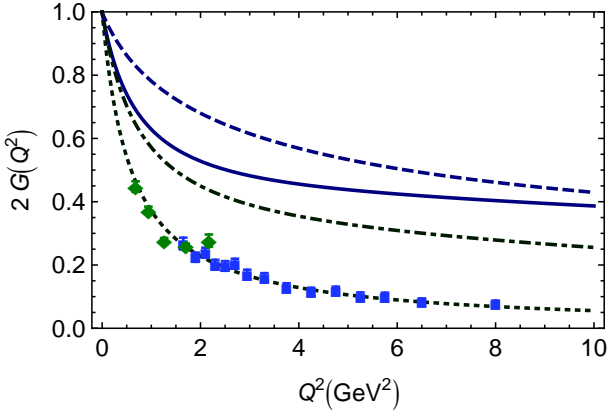}}
\caption{\label{transitionF} (Colour online)
$\gamma^\ast \pi^0 \gamma$ transition form factor in Eq.\,(\protect\ref{anomalytriangle}).
\emph{Solid curve} -- Full computation described in text, including $F_\pi(P)$ contribution;
\emph{dashed curve} -- result obtained without dressing the quark-photon vertex;
\emph{dash-dotted curve} -- result obtained with $F_\pi(P)\equiv 0$, i.e.\ forced artificially to vanish;
\emph{dotted curve} -- monopole fit to the QCD-based result in Ref.\,\protect\cite{Maris:2002mz}, which agrees with the data reported in Refs.\,\protect\cite{Behrend:1990sr,Gronberg:1997fj}, green diamonds and blue squares, respectively.
}
\end{figure}

In Fig.\,\ref{transitionF} we depict the result produced from Eq.\,(\ref{njlgluon}) using the regularisations just described.  Comparing the solid- and dashed-curves, it is evident that the effect of dressing the quark-gluon vertex diminishes with increasing $Q^2$ and therefore has no impact on the asymptotic behaviour of the transition form factor.  It does, however, affect the neutral-pion interaction-radius, which can be defined via
\begin{equation}
r_{\pi^0}^{\ast 2} = -6 \left. \frac{d}{dQ^2} \ln G(Q^2,-(m_\pi^2+Q^2)/2,0)\right|_{Q^2=0}\,.
\end{equation}
This yields $r_{\pi^0}^\ast = 0.30\,$fm and $r_{\pi^0}^{\ast\rho} = 0.45\,$fm, values that are not sensibly distinguishable from the charged-pion values listed in Table~\ref{resultsIR}.  Near equality of $r_{\pi^+}$ and $r_{\pi^0}^\ast$ is also found in the QCD-based calculations of Refs.\,\cite{Maris:2002mz,Maris:1999bh}.\footnote{Choosing instead the $\gamma^\ast \gamma^\ast \to \pi^0$ form factor, one finds $r_{\pi^0}^{\ast\ast}=0.44\,$fm and $r_{\pi^0}^{\ast\ast\rho}=0.55\,$fm, values which are larger because the momentum-scale $Q^2$ enters into both quark-photon vertices.}

More significantly, the solid- and dashed-curves in Fig.\,\ref{transitionF} show that, as with the elastic form factor \cite{GutierrezGuerrero:2010md}, the presence of the pion's necessarily-nonzero pseudovector component, $F_\pi(P)$, leads to
\begin{equation}
\label{GUVfull}
\lim_{Q^2\to \infty} G(Q^2,-(m_\pi^2+Q^2)/2,0) = {\rm constant}.
\end{equation}
This is consistent with the picture developed in Ref.\,\cite{GutierrezGuerrero:2010md}; namely, it is possible to treat the contact interaction, Eq.\,(\ref{njlgluon}), so that it yields static properties of the pion in agreement with experiment and computations based on well-defined and systematically improvable truncations of QCD's DSEs.  However, a marked deviation from experiment occurs in processes that probe the pion with $Q^2\gtrsim M^2$ and it is impossible to obtain results which agree with perturbative-QCD, even at the gross level of form-factor power-laws.

\begin{figure}[t] 
\centerline{\includegraphics[clip,width=0.47\textwidth]{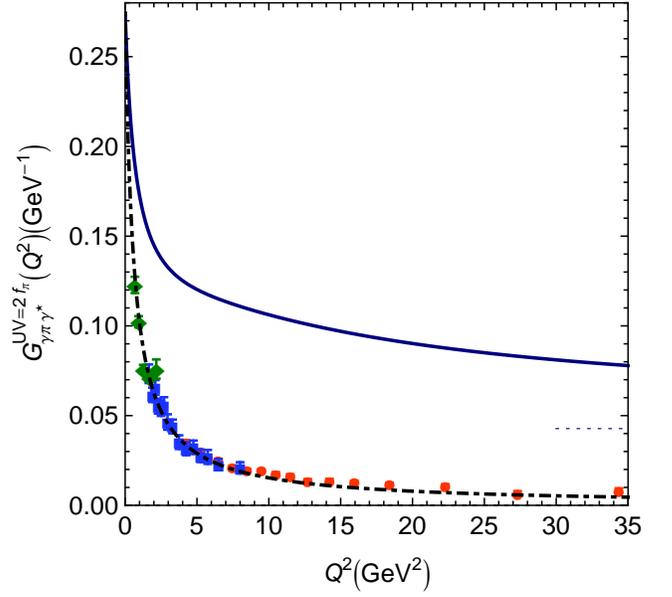}}
\caption{\label{transitionFOO} (Colour online)
$\gamma^\ast \gamma \to \pi^0$ transition form factor.
Data: red circles, Ref.\,\protect\cite{Aubert:2009mc}; green diamonds, Ref.\,\protect\cite{Behrend:1990sr}; and blue squares, Ref.\protect\cite{Gronberg:1997fj}.
\emph{Solid curve} -- $G(Q^2,0)$ computed using the symmetry-preserving, fully-self-consistent rainbow-ladder treatment of the contact interaction in Eq.\,(\protect\ref{njlgluon}), with the \emph{dotted-curve} at right showing its nonzero asymptotic limit;
and \emph{dot-dashed curve} -- fit to the $\gamma^\ast\gamma \to \pi^0$ transition form factor computed in a QCD-based rainbow-ladder-truncation DSE study \cite{Maris:2002mz}.
The curves have been divided by $(2\pi^2 f_\pi)$ in order to match the normalisation of the data.
}
\end{figure}

\begin{figure}[t] 
\centerline{\includegraphics[clip,width=0.47\textwidth]{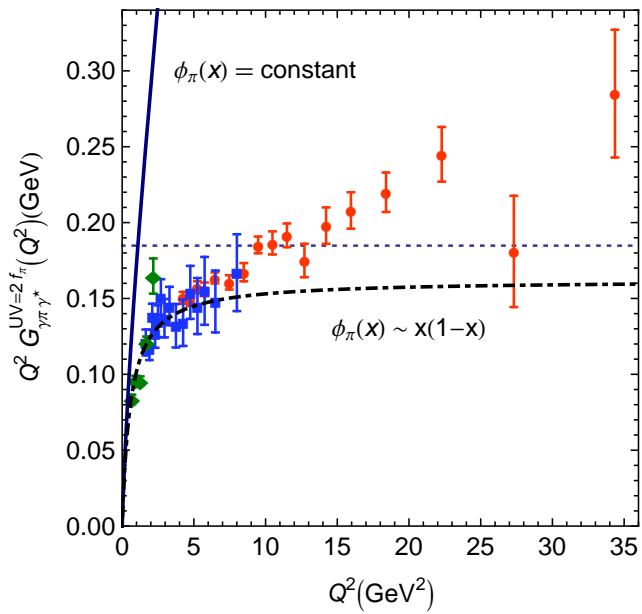}}
\caption{\label{transitionFO} (Colour online)
$Q^2$-weighted $\gamma^\ast \gamma \to \pi^0$ transition form factor.
Data: red circles, Ref.\,\protect\cite{Aubert:2009mc}; green diamonds, Ref.\,\protect\cite{Behrend:1990sr}; and blue squares, Ref.\protect\cite{Gronberg:1997fj}.
\emph{Solid curve} -- $Q^2 G(Q^2,0)$ computed using the symmetry-preserving, fully-self-consistent rainbow-ladder treatment of the contact interaction in Eq.\,(\protect\ref{njlgluon}), which produces $\phi_\pi(x)=\,$constant (see Sec.\,\protect\ref{sec:pionDA};
and \emph{dot-dashed curve} -- fit to the $\gamma^\ast\gamma \to \pi^0$ transition form factor computed in a QCD-based rainbow-ladder-truncation DSE study \cite{Maris:2002mz}.
Both curves have been divided by $(2\pi^2 f_\pi)$ in order to match the data's normalisation.
}
\end{figure}

These observations are emphasised by the comparisons presented in Figs.\,\ref{transitionFOO} and \ref{transitionFO}.  The $\gamma^\ast\gamma \to \pi^0$ form factor obtained using the symmetry-preserving, fully-self-consistent rainbow-ladder treatment of the contact interaction in Eq.\,(\protect\ref{njlgluon}) is in glaring disagreement with all existing data.  This is what it means to have a pointlike-component in the pion: all form factors must asymptotically approach a constant.  That limit rapidly becomes apparent with increasing momentum transfer because the dynamically generated mass-scale associated with low-energy hadron phenomena is $M\sim 0.4\,$GeV.  No study that neglects the pion's pseudovector component can provide a valid explanation or interpretation of the $\gamma^\ast\gamma \to \pi^0$ transition form factor, or any other of the pion's form factors.

On the other hand, it is noteworthy that the DSE result \cite{Maris:2002mz}, which is based on an interaction that preserves the one-loop renormalisation group behavior of QCD, agrees with all but the large-$Q^2$ BaBar data.

\subsection{Pion distribution amplitude}
\label{sec:pionDA}
It is worthwhile to consider a little further the nature of a pointlike pion.  As explained in Ref.\,\cite{Holt:2010vj}, with the dressed-quark propagator and pion Bethe-Salpeter amplitude in hand, one can compute the pion's valence-quark parton distribution function in rainbow-ladder truncation.  For the contact interaction, the result is
\begin{eqnarray}
\nonumber
\lefteqn{q_V(x)=  \frac{3}{2 i}{\rm tr}_{\rm D}\! \int\frac{d^4 \ell}{(2\pi)^4} \delta(n \cdot \ell - x \, n\cdot P)}\\
&\times& \Gamma_\pi(-P)\, S(\ell)\, n\cdot \gamma\, S(\ell)\, \Gamma_\pi(P)\, S(\ell-P)\,,
\label{qvxdefine}
\end{eqnarray}
where $n^2=0$, $n\cdot P=P^+$, and $x$ is the Bjorken variable.

It follows from this expression that
\begin{eqnarray}
\nonumber
\lefteqn{(n\cdot P)^{n+1} \int_0^1 dx \, x^n q_V(x) = \frac{3}{2i}
{\rm tr}_{\rm D}\! \int\frac{d^4 \ell}{(2\pi)^4} (n\cdot \ell)^n} \\
&\times& \Gamma_\pi(-P)\, S(\ell)\, n\cdot\gamma \,S(\ell)\, \Gamma_\pi(P)\, S(\ell-P)\,.
\end{eqnarray}
At this point we'll specialise to the chiral limit and: evaluate the Dirac-trace; use a Feynman parametrisation to re-express the product $\sigma_V(\ell^2)\sigma_V((\ell-P)^2)$ which arises; shift variables $\ell \to (\ell + \alpha P)$, where $\alpha$ is the Feynman parameter; use the $O(4)$ invariance of the measure to evaluate the angular integrals; and thereby arrive at
\begin{eqnarray}
\nonumber
\int_0^1 dx\, x^n q_V(x) & = & \frac{1}{n+1} \,
\frac{3}{4\pi^2} \, \overline{\cal C}_1^{\rm iu}(M^2)
E_\pi [ E_\pi - 2 F_\pi] \\
& = & \frac{1}{n+1} \,, \label{qvmoments}
\end{eqnarray}
where the last line follows because the pion's Bethe-Salpeter amplitude is canonically normalised, Eq.\,(\ref{Norm0}).

The distribution function is readily reconstructed from Eq.(\ref{qvmoments}); and one finds that even with inclusion of the pion's necessarily-nonzero pseudovector component, the contact-interaction produces
\begin{equation}
\label{qvalence}
q_V(x) = \theta(x) \theta(1-x)\,,
\end{equation}
which corresponds to a pion distribution amplitude
\begin{equation}
\phi_\pi(x)=\,\mbox{constant.}
\end{equation}
This outcome provides another way of understanding the inability of the contact interaction to reproduce the results of QCD.

As reviewed and explained in Ref.\,\cite{Holt:2010vj}, Goldstone's theorem in QCD is expressed in a remarkable correspondence between the quark-propagator and the pion's Bethe-Salpeter amplitude; i.e., between the one- and two-body problems \cite{Maris:1997hd}.  The long-known fact that the dressed-quark mass function behaves as \cite{Lane:1974he,Politzer:1976tv,Bhagwat:2003vw,Bhagwat:2006tu,Bowman:2005vx}
\begin{equation}
M(p^2) \stackrel{\mbox{large-$p^2$}}{\sim} \frac{1}{p^2}\,,
\end{equation}
entails that in QCD every scalar function in the pion's Bethe-Salpeter amplitude, Eq.\,(\ref{genpibsa}), depends on the relative momentum, $k$, as $\sim 1/k^2$ for large-$k^2$ (with additional logarithmic suppression).  It is impossible to find a kinematic arrangement of the dressed-quarks constituting the pion in which the Bethe-Salpeter amplitude remains nonzero in the limit of \mbox{infinite} relative momentum.

It follows that in QCD the pion's valence-quark distribution behaves as $(1-x)^{2+\gamma}$, $0<\gamma\ll 1$, for $x\sim 1$, at a renormalisation scale of $1\,$GeV \cite{Holt:2010vj,Hecht:2000xa}.  Hence there is no renormalisation scale in the application of perturbative QCD at which Eq.\,(\ref{qvalence}) is a valid representation of nonperturbative QCD dynamics; namely, no scale at which it is tenable to employ $\phi_\pi(x)=\,$constant, or even, more weakly, flat and nonzero at $x=0,1$.

\section{Reflections on extant data}
\label{sec:reflect}
We have shown that an internally consistent treatment of the contact interaction is incompatible with extant pion elastic (Fig.\,\ref{figrhoFpi}) and transition form factor data (Figs.\,\ref{transitionFOO}, \ref{transitionFO}).  Notwithstanding this, the results elucidated can be used in combination with QCD-based DSE studies in order to comment on available data for the process $\gamma^\ast \gamma \to \pi^0$.

\begin{figure}[t] 
\centerline{\includegraphics[clip,width=0.47\textwidth]{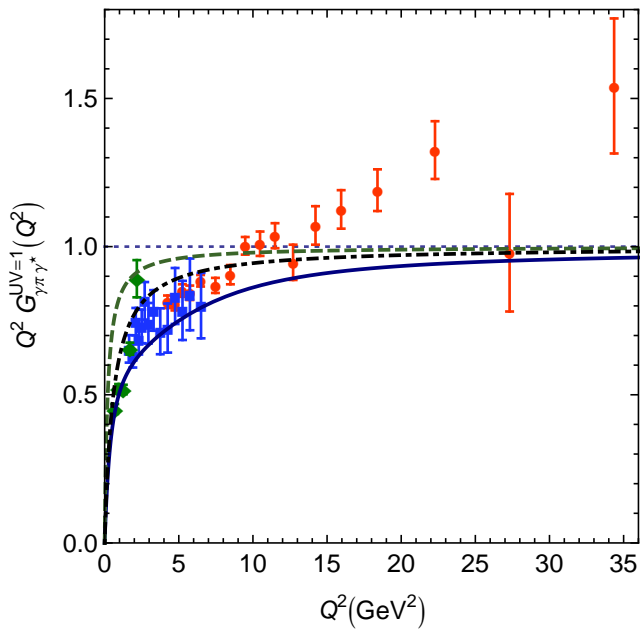}}
\caption{\label{transitionF2} (Colour online)
$Q^2$-weighted $\gamma^\ast \gamma \to \pi^0$ transition form factor.
Data: red circles, Ref.\,\protect\cite{Aubert:2009mc}; green diamonds, Ref.\,\protect\cite{Behrend:1990sr}; and blue squares, Ref.\protect\cite{Gronberg:1997fj} -- all normalised by the asymptotic form in Eq.\,(\protect\ref{BLuv}) so that the pQCD limit is marked by the dotted line at ``1''.
\emph{Solid curve} -- $Q^2 G_E(Q^2,0)$ calculated as described herein, normalised by the asymptotic form in Eq.\,(\protect\ref{GEUV});
\emph{dashed curve} -- monopole with mass-scale $(2/3) 4\pi^2 f_\pi^2$, which bounds from above the $\gamma^\ast \gamma^\ast \to \pi^0$ transition form factor computed in Ref.\,\protect\cite{Maris:2002mz}, normalised by the asymptotic form in Eq.\,(\protect\ref{gstgstuv});
and \emph{dot-dashed curve} -- monopole with mass-scale $(7/8) 4\pi^2 f_\pi^2$, obtained as fit to the $\gamma^\ast\gamma \to \pi^0$ form factor computed in Ref.\,\protect\cite{Maris:2002mz}, normalised by this same mass-scale.
}
\end{figure}

To begin we remark upon a similarity between the $Q^2$-dependence of the dash-dot- and dotted-curves in Fig.\,\ref{transitionF}; i.e., the QCD-based DSE result and that obtained from the contact interaction if the pion's pseudovector component is artificially eliminated. We emphasise that if $F_\pi(P)$ is forced to zero, then one is no longer representing faithfully the features and consequences of a vector-vector contact interaction.  Hitherto, this has nevertheless been a conventional mistreatment of Eq.\,(\ref{njlgluon}).  Its consequences were first elucidated in Ref.\protect\cite{GutierrezGuerrero:2010md}.  We describe results obtained through this intervention in order to clarify its real implications for the $\gamma^\ast \gamma \to \pi^0$ transition form factor.

An analysis of Eq.\,(\ref{anomalytriangle}) shows that the contact interaction yields
\begin{equation}
\label{GEUV}
G_E(Q^2,-(Q^2+m_\pi^2)/2,0) \stackrel{Q^2\gg M^2}{=} \sfrac{1}{2} \frac{M^2}{Q^2} \left[\ln \frac{Q^2}{M^2}\right]^2,
\end{equation}
cf.\ Eq.\,(\ref{gstgstuv}).  A similar result is obtained for the doubly-off-shell process, with the only difference being that the power of the logarithm is reduced ``$2\to 1$''.

We stress that in Eq.\,(\ref{GEUV}), $M$ is the dressed-quark mass.  Table~\ref{resultsIR} emphasises that $M$ is a computed quantity, which is completely determined once the interaction and truncation are specified.  The value of $M$ is tightly connected with those of all other measurable properties in the Table.  Thus, in a well-constrained and internally-consistent analysis, one cannot significantly alter $M$ without materially changing all the other static properties which characterise the pion.  No theoretical analysis is reliable if it allows itself to skirt these constraints.

In Fig.\,\ref{transitionF2} we depict the contact-interaction result for $Q^2 G_E(Q^2,-(Q^2+m_\pi^2)/2,0)$ normalised by the asymptotic form in Eq.\,(\ref{GEUV}).  In addition, we plot a monopole with mass-scale $(2/3) 4\pi^2 f_\pi^2$, which bounds uniformly from above the QCD-based DSE calculation of the $\gamma^\ast\gamma^\ast \to \pi^0$ transition form factor reported in Ref.\,\cite{Maris:2002mz}; and a monopole with mass-scale $(7/8) 4\pi^2 f_\pi^2$, which is a fit to the $\gamma^\ast\gamma \to \pi^0$ transition form factor also computed therein.  The origin of these mass-scales was discussed in Sec.\,\ref{UVlimits}.  It is striking that these curves all approach their asymptotic limits from below.  Stated differently, each is a monotonically-increasing concave function.  Indeed, this is even true of the solid curve in Fig.\,\ref{transitionFO}.

\section{Conclusions}
\label{sec:end}
We have shown that in fully-self-consistent treatments of pion: static properties; and elastic and transition form factors, the asymptotic limit of the product $Q^2 G_{\gamma^\ast\gamma \pi^0}(Q^2)$, which is determined \emph{a priori} by the interaction employed, is not exceeded at any finite value of spacelike momentum transfer: the product is a monotonically-increasing concave function.  We understand a consistent approach to be one in which: a given quark-quark scattering kernel is specified and solved in a well-defined, symmetry-preserving truncation scheme; the interaction's parameter(s) are fixed by requiring a uniformly good description of the pion's static properties; and relationships between computed quantities are faithfully maintained.

Within such an approach it is nevertheless possible for $Q^2 F^{\rm em}_\pi(Q^2)$, with $F^{\rm em}_\pi(Q^2)$ being the elastic form factor, to exceed its asymptotic limit because the leading-order matrix-element involves two Bethe-Salpeter amplitudes.  This permits an interference between dynamically-generated infrared mass-scales in the computation.  Moreover, for $F^{\rm em}_\pi(Q^2)$ the perturbative QCD limit is more than an order-of-magnitude smaller than $m_\rho^2$.  Owing to the proximity of the $\rho$-meson pole to $Q^2=0$, the latter mass-scale must provide a fair first-estimate for the small-$Q^2$ evolution of $F^{\rm em}_\pi(Q^2)$.  A monopole based on this mass-scale exceeds the pQCD limit $\forall Q^2>0$.  For the transition form factor, however, the opposite is true because $m_\rho^2$ is less-than the pQCD limit, Eq.\,(\ref{BLuv}).

A vector current-current contact-interaction may be described as a vector-boson exchange theory with vector-field propagator $(1/k^2)^\kappa$, $\kappa=0$.  We have shown (see Figs.\,\ref{transitionFOO}, \ref{transitionFO}) that the consistent treatment of such an interaction produces a $\gamma^\ast\gamma \to \pi^0$ transition form factor that disagrees with \emph{all} available data.  On the other hand, precisely the same treatment of an interaction which preserves the one-loop renormalisation group behaviour of QCD, produces a form factor in good agreement with all but the large-$Q^2$ data from the BaBar Collaboration \cite{Aubert:2009mc}.

Studies exist which interpret the BaBar data as an indication that the pion's distribution amplitude, $\phi_\pi(x)$, deviates dramatically from its QCD asymptotic form, indeed, that $\phi_\pi(x)=\,$constant, or is at least flat and nonvanishing at $x=0,1$ \cite{Radyushkin:2009zg,Polyakov:2009je}.  We have explained that such a distribution amplitude characterises an essentially-pointlike pion; and shown that, when used in a fully-consistent treatment, it produces results for pion elastic and transition form factors that are in striking disagreement with experiment.  A bound-state pion with a pointlike component will produce the hardest possible form factors; i.e., form factors which become constant at large-$Q^2$.

On the other hand, QCD-based studies produce soft pions, a valence-quark distribution amplitude for the pion that vanishes as $\sim (1-x)^2$ for $x\sim 1$, and results that agree well with the bulk of existing data.

Our analysis shows that the large-$Q^2$ BaBar data is inconsistent with QCD and also inconsistent with a vector current-current contact interaction.  It supports a conclusion that the large-$Q^2$ data reported by BaBar is not a true representation of the $\gamma^\ast\gamma \to \pi^0$ transition form factor, a perspective also developed elsewhere \cite{Mikhailov:2009sa}.  We are confirmed in this view by the fact that the $\gamma^\ast \to \eta \gamma$ and $\gamma^\ast \to \eta^\prime \gamma$ transition form factors have also been measured by the BaBar Collaboration \cite{Aubert:2006cy}, at $Q^2=112\,$GeV$^2$, and in these cases the results from CLEO \cite{Gronberg:1997fj} and BaBar are fully consistent with perturbative-QCD expectations.




\begin{acknowledgments}
We acknowledge valuable: correspondence with S.\,J.~Brodsky; and discussions with C.~Hanhart, R.\,J.~Holt and S.\,M.~Schmidt.
This work was supported by:
Forschungszentrum J\"ulich GmbH;
the U.\,S.\ Department of Energy, Office of Nuclear Physics, contract no.~DE-AC02-06CH11357;
the Department of Energy's Science Undergraduate Laboratory Internship programme;
CIC and CONACyT grants, under project nos.\ 4.10 and 46614-I;
and the U.\,S.\ National Science Foundation under grant no.\ PHY-0903991 in conjunction with a CONACyT Mexico-USA collaboration grant.
\end{acknowledgments}

\end{document}